\documentclass{IEEEtran}

\ifCLASSINFOpdf
  \usepackage[pdftex]{graphicx}
  \graphicspath{./imgs/}
  \DeclareGraphicsExtensions{.pdf,.jpeg,.png}
\else
  \usepackage[dvips]{graphicx}
  \graphicspath{./imgs/}
  \DeclareGraphicsExtensions{.eps}
\fi

\usepackage{epstopdf}
\usepackage{tikz}
\usepackage{url}
\usepackage[cmex10]{amsmath}
\usepackage{mathtools} % Para tener los enviroments con *
\usepackage{color}
\usepackage{cleveref}

\usepackage[ruled,vlined]{algorithm2e}

\usepackage{cite}

\usepackage[export]{adjustbox}

\ifCLASSOPTIONcompsoc
  \usepackage[caption=false,font=normalsize,labelfont=sf,textfont=sf,justification=centering]{subfig}
\else
  \usepackage[caption=false,font=footnotesize,justification=centering]{subfig}
\fi
\usepackage{stfloats}
\usepackage{algpseudocode}
\usepackage{amsfonts}
\usepackage{float}
\usepackage{multirow}

\usepackage[colorinlistoftodos]{todonotes}

\algnewcommand\algorithmicinput{\textbf{Input:}}
\algnewcommand\Input{\item[\algorithmicinput]}
\algnewcommand\algorithmicoutput{\textbf{Output:}}
\algnewcommand\Output{\item[\algorithmicoutput]}

\begin{document}
\title{Location Method for Forced Oscillation Sources Caused by Synchronous Generators}

\author{Pablo Marchi, Pablo Gill Estevez, and Cecilia Galarza 
\thanks{P. Gill Estevez, P. Marchi and C. G. Galarza work with the School of Engineering, Universidad de Buenos Aires and the CSC-CONICET, Argentina. (e-mail: pgill@fi.uba.ar, pmarchi@fi.uba.ar, cgalar@fi.uba.ar)}}%
\maketitle

\begin{abstract}
In this article, we present a new methodology to identify if forced oscillation sources are generated in excitation systems or in turbine governors. In this context, we propose to harness the information that the dynamic state estimation (DSE) provide 
to be able to apply a dissipating energy flow (DEF) method. The measurements that the phasor measurement units (PMUs) provide are used to estimate the internal states of different generators connected to the same bus. Here, it will be considered that the PMU at the point of connection is only capable of measuring the voltage phasor and the total current phasor. Therefore, the current injection of the generator that is causing the FO is an unobservable variable. To overcome this issue, event playback is applied to simulate the model's response and a comparison between the DSE of multiple generators is made. Finally, when the faulty generator is identified, we compute energy functions for the mechanical control loop and for the excitation control loop. The sign and rate of these energies determine the source of the oscillation, allowing that the FO can be located. Simulated signals are used to show that the proposed method 
provides a systematic methodology for identification and location of power systems forced oscillations than even could be 
non-stationary signals.

\end{abstract}

\begin{IEEEkeywords}
Forced oscillations, phasor measurement unit (PMU), dynamic state estimation, Unscented Kalman filter, non-stationary signal. 
\end{IEEEkeywords}

% \fi
\IEEEpeerreviewmaketitle

\section{Introduction}

\IEEEPARstart{C}{ontrary} to free oscillations, which mainly depend on the dynamic characteristics of the system, forced 
oscillations (FOs) are determined by external inputs and disturbances that periodically excite the power system 
\cite{GHORBANIPARVAR2017}. FOs can occur in power systems due to different reasons, such as inadequate control designs, 
equipment failure and abnormal generator operating conditions \cite{Nerc2017}. The identification of FOs is crucial because the
resonance between FOs and the electromechanical modes of the system can cause system break-down \cite{Vournas1991}. The most
effective countermeasure is separating the external disturbance from the system, but the FO should be located first. Usually
the disturbance is affecting a generator, and the most common measure is tripping the generator \cite{Chen2013}. However,
generator owners and grid operators could be interested in avoiding unnecessary disconnections. They would like to be capable of mitigating the
oscillations before severe failures in the equipment or critical disturbances on the grid take place. If the disconnection can not be avoided, then it would be
interesting to identified at least which control loop of the machine is causing the abnormal state of operation before the
disconnection occurs.   

\begin{figure}[]
\centering
\includegraphics[width = .9\columnwidth]{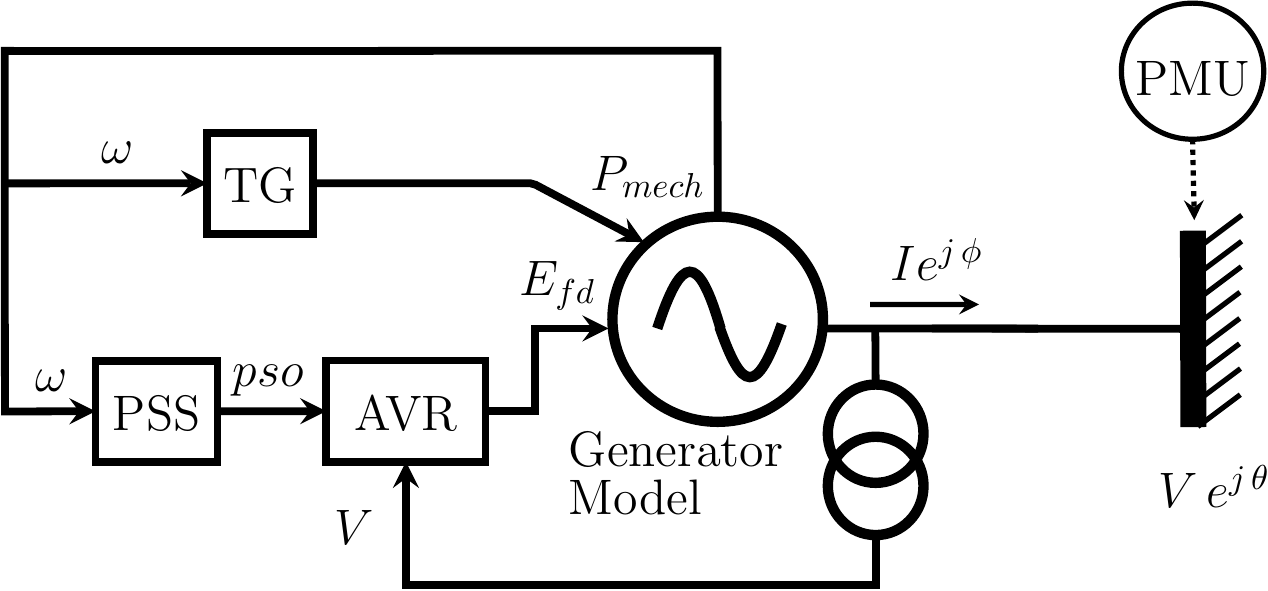}
\caption{Generator and PMU at the point of connection.}
\label{fig:Sec2:gen_controllers}
\end{figure}

Fig. \ref{fig:Sec2:gen_controllers} shows the general structure of a power plant. The system 
consists of a synchronous generator, a turbine governor (TG), an automatic voltage regulator (AVR) and a power system 
stabilizer (PSS). Here, we consider that a phasor measurement unit (PMU) is installed at the connection point and that it provides high quality magnitude and phase measurements to monitor the generator unit. In spite of the system is being monitored, one of these control loops can introduce a FO and not being identified. Many methods for locating the source have been proposed in the past few years \cite{Dan2018}, among these methods, 
Dissipating Energy Flow (DEF) method \cite{Chen2013} has shown a good performance for oscillation source location, and it
was recently adopted by Independent System Operator - New England (ISO-NE) \cite{MASLENNIKOV201755}. In this system, non-sinusoidal and non-stationary nature of the oscillation have been identified.

On the other hand, dynamic state estimation (DSE) is a new power system functionality that is capable of accurately capturing the dynamics of the system states, and it will play an important role in power system 
control and protection \cite{Zhao2019_review}. The main objective of DSE is to obtain a real time estimation of the state of the power system. One way to achieve this 
goal is to implement a Bayesian estimation algorithms \cite{Zhou2015_bis, Aghamolki2015, Zhao2019}. The implementation of these algorithms is based on a state-space model, 
where a transition and measurement functions of the system are defined. However, to
estimate the state vector even under abnormal operating conditions is hard to accomplish because all the possible internal faults should be modeled. As the possible number of models to assign to a FO is huge, it would not be advisable to assume a model for the TG, PSS and AVR. For this reason, a sigma-point based Kalman filtering with unknown inputs is used \cite{Anagnostou2018}. This 
solution provides us a tool capable of estimating the state variables of the system even if the dynamics of some of them 
are not defined. For example, if the mechanical power ($P_{mech}$) and the field voltage ($E_{ef}$) are considered as an unknown inputs, the equations of 
the TG, the PSS and the AVR are not incorporated in the transition function of the model, and the variables $P_{mech}$ and $E_{ef}$ can still be estimated. Unfortunately, it is only 
possible under certain conditions of observability of the system.

The main goal of this paper is to develop a systematic methodology to identify which generator is causing the FO, even if it is
connected to a bar with other generators. After that, the identification of the control loop associated with synchronous generator is required. As the main contribution, we introduce a new procedure which uses the information 
that the DSE techniques are capable of obtaining and we combine it with a DFE method to find the oscillation source. 
Finally, we evaluate the performance of our proposal on simulated data considering different scenarios.

\section{Mathematical Background}

\subsection{Dynamic state estimation with unknown inputs} \label{sec:DSE}

To estimate the dynamics of a generator, and its control systems, the following general form of a discrete-time state-space model
for nonlinear systems is considered:
\begin{align}
\mathbf{x}_{k} &= f_s(\mathbf{x}_{k-1},\mathbf{u}_{k-1}) + G_k \, \mathbf{d}_{k-1} +\mathbf{w}_{k}, \label{eq:nonlinearsystem1}\\
\mathbf{z}_{k} &= h_s(\mathbf{x}_{k},\mathbf{u}_{k})+\mathbf{v}_{k}. \label{eq:nonlinearsystem2}
\end{align}

where $f_s$ is the state transition function which models the system dynamics, $\mathbf{x}\in\mathbb{R}^{n \times 1}$ is 
the state vector, $\mathbf{z}\in\mathbb{R}^{m \times 1}$ is the measurement vector, $\mathbf{u}\in\mathbb{R}^{l \times 
1}$ is the input vector, $h_s$ is the measurement function which relates the measurements with the state vector, $G\in\mathbb{R}^{n \times 
p}$ is the unknown 
input distribution matrix, $\mathbf{d}\in\mathbb{R}^{p \times 
1}$ is the unknown input vector and 
$\mathbf{w}\in\mathbb{R}^{n \times 1}$ and $\mathbf{v}\in\mathbb{R}^{m \times 1}$ are noise vectors introduced to account
for modeling errors for each time step $k$.
 
Given its simplicity, its reduced computational cost, and its good performance for non-linear systems, the unscented Kalman 
filter (UKF) is one of the algorithms most used in the literature to obtain an estimator for the state vector for each time 
frame. Indeed, as it was shown in \cite{Zhou2015_bis}, the UKF is a feasible real time solution. This algorithm defines the unscented 
transformation to approximate
the mean $\mathbf{x}_k$ and covariance $P_k$ of the state vector. For this purpose, the concept of sigma points 
is introduced. These sigma points are propagated through the nonlinear functions. Then, the mean and the covariance for 
$\hat{\mathbf{x}}$ and $\hat{\mathbf{z}}$ are approximated using 
a weighted sample mean and covariance of the posterior sigma points. Basically, the procedure can be divided in two 
stages: prediction and correction. 

A new variant of the UKF is the the UKF with unknown inputs algorithm (UKF-UI), It will be 
shortly described below because it represents the core of the proposed methodology, for an in-depth discussion of the UKF-UI 
algorithm, the reader is referred to \cite{Anagnostou2018}. Here, voltage phasors, current 
phasors and frequency measurements obtained by a PMU, or other intelligent electronic device, are used to define the input 
vector ($\mathbf{u}$) and as the measurement vector ($\mathbf{z}$) in the model described by equation 
\eqref{eq:nonlinearsystem1} and \eqref{eq:nonlinearsystem2}. The complete process can be described performing the following three steps:

\begin{itemize}
\item Biased State Estimation: In this step, it is considered that there is no prior information regarding the
unknown input of the previous step. So, the unknown input vector is set to zero. Then, an estimation of 
the state ($\hat{\mathbf{x}}_k^b$), the covariance matrix of the state vector ($P_k^b$), a measurement 
prediction ($\hat{\mathbf{z}}_k^b$) and the cross-covariance between the predicted states and measurements ($P_{xz\,k}^b$) are obtained using the same equations established by the UKF. 
\item Unknown Input Estimation: A linear approximation of the measurement function is used \cite{Terejanu2007}:
\begin{equation}
    \begin{aligned}
        h_s(\mathbf{x}_k,\mathbf{u}_k) &\approx H_{k} \mathbf{x}_k + \mathbf{c}_k, \\
        H_k &= (P_{xz\,k}^b)^T (P_k^b)^{-1}, \\
        \mathbf{c}_k &= \hat{\mathbf{z}}_k^b - H_k \hat{\mathbf{x}}_k^b
    \end{aligned}
\end{equation}
Then, and a least square linear regression problem can be formulated as:
\begin{equation}
    \begin{aligned}
        \Tilde{\mathbf{z}}_k &= \mathbf{z}_k - \hat{\mathbf{z}}_k^b = H_k G_k \, \mathbf{d}_{k-1} + \Delta_k \\
        \Delta_k &= H_k \left(f_s(\mathbf{x}_{k-1},\mathbf{u}_{k-1})-\hat{\mathbf{x}}_k^b\right) + \mathbf{v}_k, 
    \end{aligned}
\end{equation}
where the equivalent noise vector $\Delta$ has the following properties:
\begin{equation}
    \begin{aligned}
        E \left[ \Delta \right] &= \mathbf{0}, \\
        E \left[\Delta_k \Delta_k^T \right] &= H_k P_k^b H_k^T + R_k = \Tilde{R}_k.
    \end{aligned}
\end{equation}
Here, $R \in \mathbb{R}^{m \times m}$ is the covariance matrix of the measurement noise. Finally, a least square estimator for the unknown input can be defined using the covariance matrix of the resulting noise and the pseudoinverse of the $H_{k} G_k$:
\begin{equation}
    \hat{\mathbf{d}}_{k-1}=\left(G_k^T H_{k}^T \Tilde{R}_k^{-1} H_{k} G_k \right)^{-1} G_k^T H_{k}^T \Tilde{R}_k^{-1} \Tilde{\mathbf{z}}_k
\end{equation}
\item Unbiased State Estimation: After obtain an estimation of the unknown vector, the equations of the UKF algorithm are applied again. 
Here, a new set of sigma points is calculated considering the model described in \eqref{eq:nonlinearsystem1} and \eqref{eq:nonlinearsystem2}. An estimation of 
the state ($\hat{\mathbf{x}}_k^{u-}$), the covariance matrix of the state vector ($P_k^{u-}$), a measurement 
prediction ($\hat{\mathbf{z}}_k^u$), a predicted measurement covariance ($P_{z\,k}^u$) and the cross-covariance between the predicted states and measurements ($P_{xz\,k}^u$) for this unbiased condition are obtained. Then, they are used to compute the final estimation of the state vector ($\hat{\mathbf{x}}_k^{u+}$) and its covariance matrix ($P_k^{u+}$), using the Kalman filter gain ($K$) and the residual or innovation vector ($\mathbf{y}$) as follows:  
\begin{equation}
    \begin{aligned}
        \hat{\mathbf{x}}_k^{u+} &= \hat{\mathbf{x}}_k^{u-} + K_k  \mathbf{y}_k , \\
        K_k &= P_{xz\,k}^u\,(P_{z\,k}^u)^{-1}, \\
        \mathbf{y}_k &= \mathbf{z}_k - \hat{\mathbf{z}}_k^u, \\
        P_k^{u+} &= P_k^{u-} - K_k P_{xz\,k}^u K_k^T
    \end{aligned}
    \label{eq:UKF_UI_unbias}
\end{equation}
\end{itemize}

\subsection{Dissipating Energy Flow Method} \label{sec:DEF}

Oscillatory signals can be divided in different components and the damping of a each component can be estimated by its energy 
dissipation. However, constructing energy functions for various models is a difficult problem. An energy function construction 
method was developed in \cite{Min2007}, leading to the energy function obtained by \cite{Chen2013}:
\begin{equation}
W_h^{ij} \approx \int P_h^{ij} \, d\theta_h^i + \int Q_h^{ij} \frac{dV_h^i}{V^i},
\label{eq:DSE_function}
\end{equation}
where $P^i$ and $Q^i$ are the active and reactive power flows in branch $i$, $\theta^i$ is the bus voltage angle, $V^i$ is the bus voltage magnitude and index $h$ indicates the corresponding oscillatory component. The value and sign of the rate of change of $W_h^{ij}$ have a physical interpretation as the amount and direction of the dissipating energy flow. Calculated by \eqref{eq:DSE_function}, energy $W_h^{ij}$ in each branch is either increasing (positive) or decreasing (negative) over time. Negative means the dissipating energy flows into the bus from the source. Positive
means the energy flows from the bus to a system element dissipating transient energy. In other words, it indicates the direction of the source location relative to the branch $ij$. Positive rate of change of $W_h^{ij}$ means the source is located behind bus $i$, and a negative value means the source is located behind bus $j$ or branch $ij$ is the source \cite{Maslennikov2020}.

In \cite{Chen2014} a oscillation energy analysis method for the evaluation of generator damping at the point of connection is developed. The oscillation energy 
flow in a generator and the energies dissipation of the field winding and the damper winding are studied. There, the generator is represented by the dynamic equations of the fourth-order and two-axis generator model \cite{sauer2006power}. In Appendix \ref{sec:nomenclature}, a complete list for the variables and parameters used for the model is included. Based on the analysis made in \cite{Chen2014}, the oscillation energy for all the components and flowing from the generator to the system can be divided into two parts:
\begin{align}
     W_e &= \frac{1}{X_d-X'_d} \int \left( E_{fd} - T'_{d0} \Dot{E}'_q \right) dE'_q  \nonumber\\
     &- \frac{T'_{q0}}{X_q-X'_q} \int \Dot{E}'_d\,dE'_d \nonumber\\
     &- \frac{1}{2} \left( \frac{E'_q{}^2}{X_d-X'_d}+\frac{E'_d{}^2}{X_q-X'_q} + X'_d I_d^2 + X'_q I_q^2 \right), \label{eq:W_e} \\
    W_g &= \int P_{mech} d\delta - H \omega^2.  \label{eq:W_g}
\end{align}
Here, the damping factor of the generator model was neglected \mbox{($D=0$)}. Now, it must be taking into account that the 
independent integral terms from equations \eqref{eq:W_e} and \eqref{eq:W_g} are related to the transient 
energy of the generator \cite{Tsolas1985}, and the second term of \eqref{eq:W_e} is related to the energy of 
the rotor windings \cite{Chen2014}. The rest of the terms are associated to the field winding and the governor, so the functions for the control loops can be defined as:

\begin{align}
     W_{field}&= \frac{1}{X_d - X'_d} \int \left( E_{fd} - T'_{d0}\Dot{E}'_q \right) \, d E'_q \nonumber\\
     &= \frac{T'_{d0}{}^{-1}}{(X_d - X'_d) } \int X_{ad} I_{fd} \left( E_{fd} - X_{ad} I_{fd} \right) \, d t, \\
    W_{mech} &= \int P_{mech} \,d \delta
\end{align}

For discrete signals, a discrete-time approximation using the Euler method and the trapezoidal rule is performed to obtain the formulation required to compute these energies:
\begin{align}
W_{field}[k] &= W_{field}[k-1] + \frac{T'_{d0}{}^{-1}\,\Delta_t}{X_d-X'_d} \left[ E_{fd}[k] \right. \nonumber \\
& \left. \,X_{ad}I_{fd}[k]-(X_{ad}I_{fd}[k])^2 \right], \\
 W_{mech}[k] &= W_{mech}[k-1] + (P_{mech}[k]+P_{mech}[k-1])\nonumber \\
 & (\delta[k]-\delta[k-1])\frac{1}{2},
\label{eq:flow_disip_energy}
\end{align}
where $k$ represents the time instant and $\Delta_t$ the time step.

\section{Proposed Methodology} \label{sec:Method}

As it was mentioned before, the problem to analyze is to locate the source of the FO when it is unknown, and it could be any
generator connected at a specific bus. Indeed, the proposed methodology is applicable after the bus where the FO occurred
has been identified, where any variant of the DEF method could be used to this end. As an example, Fig. \ref{fig:results_test_bench} shows a sub-system
where the FO take place in generator H and the rest of them continue with normal operation. Besides, it is important to
emphasize that it is assumed that a PMU can measure one current branch only. In this example, the current flow from bus 6132
to the bus 6102 is being measured. Under these circumstances, where the the current of faulty generator is not being measured, the
lack of observability impedes that the algorithm described in section \ref{sec:DSE} can be implemented directly. To overcome this issue, new estimation methodologies are required.

\begin{figure}[]
\centering
\vspace*{-2mm}
\includegraphics[width = 0.75\columnwidth]{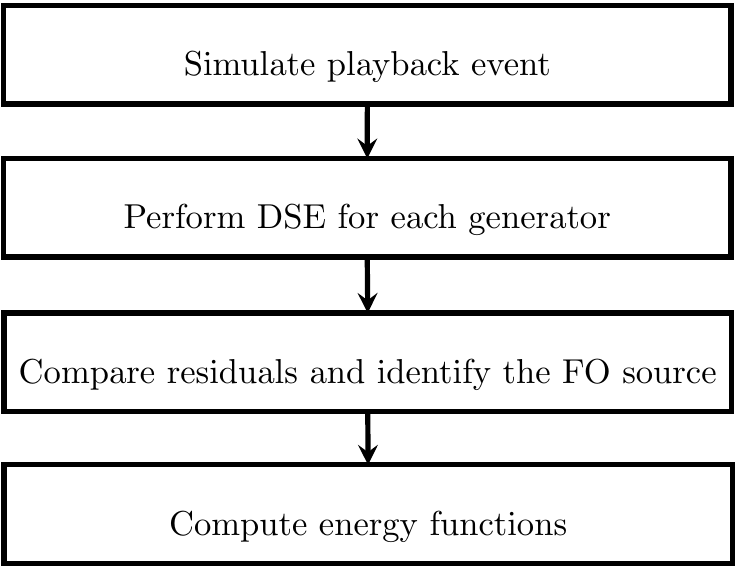}
%\vspace*{-5mm}
\caption{Flow chart of the proposed methodology.}
\vspace*{-2mm}
\label{fig:dia_ppal}
\end{figure}

Fig. \ref{fig:dia_ppal} shows the flow chart of the proposed methodology. The first step consist in performing a \textit{event 
playback} \cite{Huang2013}. This event consist in perform a dynamic evolution of the states of each generator considering the 
equation \eqref{eq:nonlinearsystem1}, but neglecting the process noise ($\mathbf{w}$) and the unknown inputs ($\mathbf{d}$). To perform the playback, the voltage magnitude $V$ and the phase at the point of connection $\theta$ are used as inputs in the models of each generator. Besides, an initial condition is required and it will be considered as the value obtained by the SCADA system, with an error that does not exceed 3\%. In this step, the use of unknown inputs is no longer required since the transition function will include the dynamics of the TG, the PSS and the AVR as the generator was operating normally. After that, the current injection phasor of generator $j$ is calculated as:
\begin{equation}
I_j = I_T - \sum_{i=1, i\neq j}^{T} I_i,
\label{eq:inyections}
\end{equation}
where $T$ is the total number of generators connected to the same bus and $I_T$ is the current phasor flowing form the corresponding
bus to the rest of the system. Obviously, when a FO occurs there will be an considerable error in the current injections 
values using the equation \eqref{eq:inyections}. It is important to mention that the error of the current associated to the
generator where the FO take place will be less than in other cases because a model considering a normal operation is used
for the other generators.

Next, we perform a DSE using the algorithm described in section \ref{sec:DSE} for each generator. The main differences with the work made in \cite{Anagnostou2018} are that the complete equations of the GENROU model are used \cite{Marchi2020}, including the saturation functions, and the following definitions for the state, measurement and input vectors are made:
\begin{equation}
\begin{aligned}
    \mathbf{x} &= \left[ \delta \,\,\, \omega \,\,\, E'_d \,\,\, E'_q \,\,\, \Psi_d' \,\,\,  \Psi_q' \right], \qquad \mathbf{d} =  \left[ P_{mech} \,\,\, E_{fd} \right]\\
    \mathbf{z} &= \left[ V_{re}\quad V_{im} \right], \qquad     \mathbf{u} = \left[ I_{re}\quad I_{im} \right]
\end{aligned}   
\end{equation}

Besides, we have considered some constrains in the state variables. They are constrained by the projection operator $\mathcal{P}$ to restrict them to the feasible region of the state-space. In this case, the operator is
defined as a saturation based on the generator model parameters. For example, the $E_{fd}$ variable will be saturated by the
values $[Efd_{max}, Efd_{min}]$ when the computation of the sigma points is made. This modification to the basic algorithm is 
known as constrained UKF (CUKF). For an in-depth discussion of the CUKF algorithm, the reader is referred to \cite{Zhao2019}. 
The parameters for the CUKF were selected using the same criteria as in \cite{Aghamolki2015}.

The following step is to compare the residuals from \mbox{equation \eqref{eq:UKF_UI_unbias}}, between the DSE results for each 
generator. The maximum residual $y_{max}$ from the residual vector $\mathbf{y}$ will be consider as the most representative 
value. After that, the energy \mbox{$E(y_{max}) = \sum_{i=0}^{L} | y_{max} [i] |^2$} is calculated, where $L$ is the total 
window length for a specific instant of time. Then, the generator identified as the source of the FO will be the one with the smaller energy. This fact is based on the assumption that  the DSE is accurately enough only if it is applied to the generator where the FO occurred, and the rest of the models are considered as normal operation. Hence, this assumption restrain that our proposal to be applicable in situations where only one FO is happening at a time.  

Finally, once the generator is identified, the energy functions described in section \ref{sec:DEF} are computed using equations \eqref{eq:flow_disip_energy}. The energy $W_{pmech}$ is related to the mechanical control loop while $W_{field}$ is related to the excitation control loop. As it was established in \cite{Maslennikov2020}, the location of the FO is identified using the sign and the rate of change of the energies. Hence, the FO will be associated to the control loop which energy function has a positive rate or tendency. 

\section{Numerical Results}
\begin{figure}[]
\centering
\vspace*{-2mm}
\includegraphics[width = 1\columnwidth]{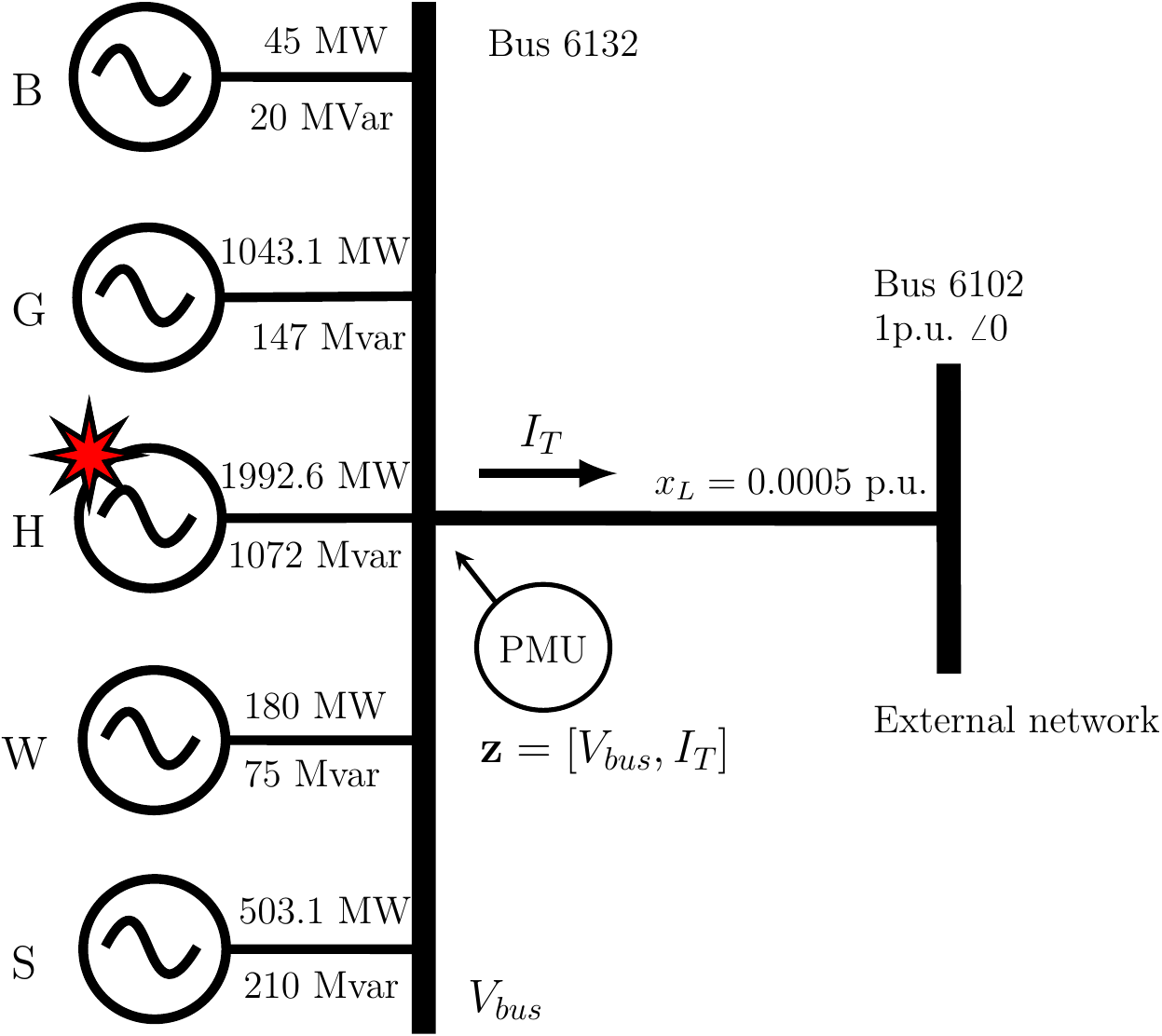}
%\vspace*{-5mm}
\caption{Test bench system with five generators. B: Biomass, G: Gas, H:  Hydro, W: Wind, S: Solar.}
\vspace*{-2mm}
\label{fig:results_test_bench}
\end{figure}

In this section, we evaluate the performance of our proposal in two different scenarios. The first scenario (A) consist 
in simulations of FOs using the test bench system presented in \mbox{Fig. \ref{fig:results_test_bench}}. This test bench system considers five generators connected to the same bus, and a FO takes place place in generator H. The second scenario (B) is based on analyzing the cases of the 2021 IEEE-NASPI Oscillation Source Location (OSL) Contest. All test cases used for this contest were generated by simulating a WECC 240-bus test system \cite{Price2011} developed by NREL based on reference \cite{Yuan2020}. The system has 243 buses, 146 generating units at 56 power plants (including 109 synchronous machines and 37 renewable generators), 329 transmission lines, 122 transformers, 7 switched shunts and 139 loads. Indeed, scenario (A) is a subsystem (bus number 6132) of scenario (B) where all the dynamic state variables of the generators will be compared with the estimates obtained by the DSE algorithm.

\subsection{Test bench system}

In this scenario, all the simulations are performed using the PSS/E software and the system presented in \mbox{Fig 
\ref{fig:results_test_bench}}. The model used for the synchronous generators is the GENROU model and the SEXS model for 
modeling the excitation system. Finally, the turbine governors are modeled using the TGOV1, GAST, HYGOV models for 
generators B, G and H respectively. All the models used can be found in the PSS/E manual \cite{PSSEModelLibrary}. The PSS
of each synchronous generator are not modeled for this scenario. Renewable machines are modeled using REGCA1 for the generator and the REECB1 model for the electrical control system. The initial conditions are specified in Fig. \mbox{Fig. \ref{fig:results_test_bench}} and they are assumed to be known.

The simulated measurements for each scenario include the voltage and current phasors, as well as the frequency at the 
point of connection (bus 6132). The simulation time step is set to $T_s = 1$ ms. Here, we are considering that the PMU 
can transmit the phasor measurement data almost at its sample rate. This is not a far-fetched assumption since it is 
assumed that the measurements will be used by the generator owner as a stand-alone application. Besides, an additive 
white Gaussian noise (AWGN) was considered for the voltage, current and frequency  measurements. Concretely, the voltage 
and current noises are modeled as a complex and circularly-symmetric random  variable, where the variance of the noise is
adjusted as $\sigma_{re}=\sigma_{im}=TVE/(3\,\sqrt(2))$. Then, the following  criterion for the total vector error (TVE) 
standard metric: $TVE = 1\%$ and a frequency error metric $FE = 0.5$ mHz are chosen. Although these values are 
considerable less (ten times) than the actual requirements of synchrophasors measurements \cite{IECStandard2018} they 
imply to be working with signals with a $SNR_{FO} = 20$dB or less, where
\begin{equation}
SNR_{FO}[dB] =20 \log_{10} (\mathrm{Var}[\mathrm{FO}_{signal}] / \sigma_v^2)
\end{equation}
is defined as the  Signal-to-Noise  Ratio between the FO signal and the noise. In the following subsections, two types of
FOs are considered, where each one is originated in different control loops.    

\vspace{0.1cm}

\subsubsection{Modulation of the field voltage $E_{fd}$}

\begin{figure}[]
\centering
\vspace*{-2mm}
\includegraphics[width = 1\columnwidth]{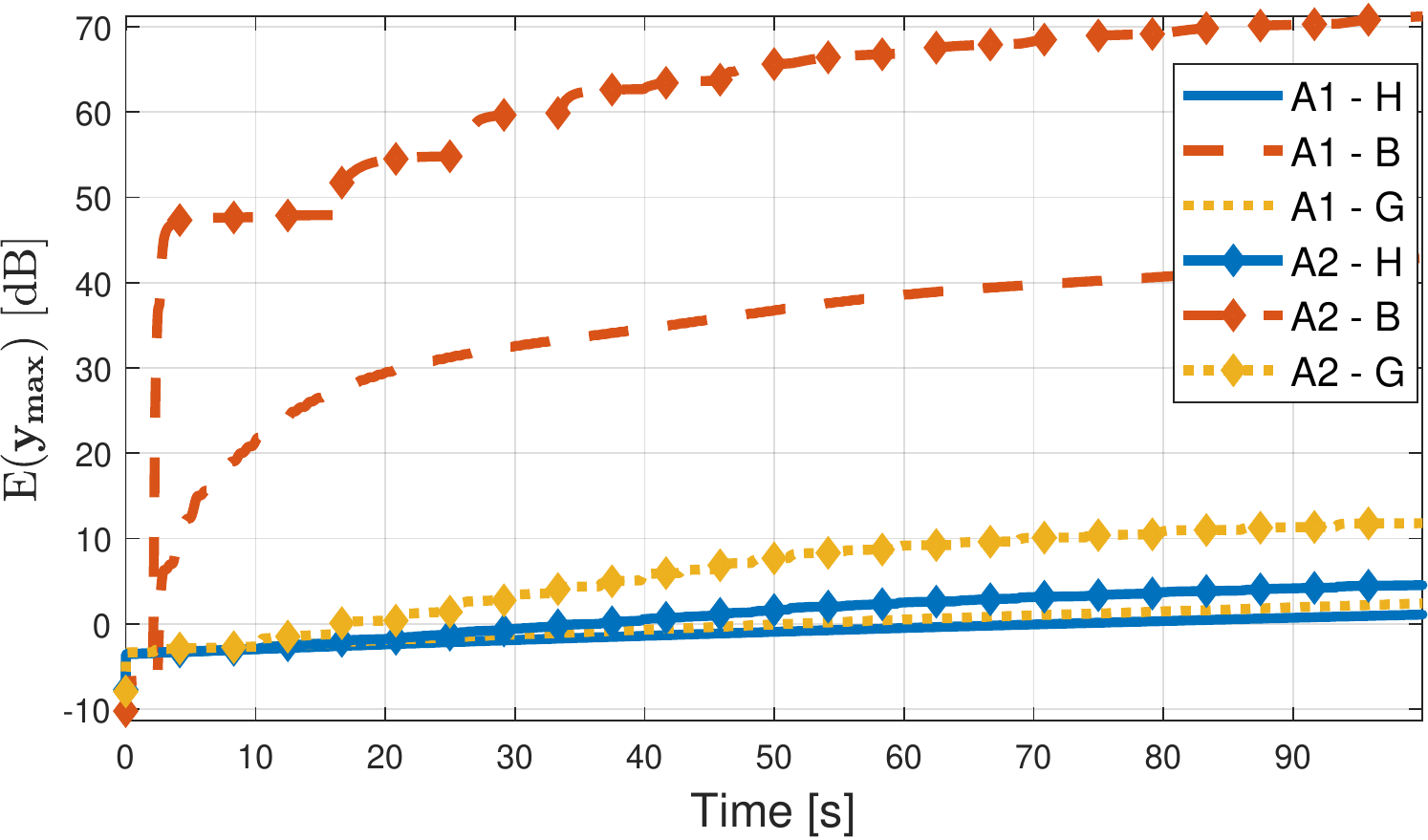}
%\vspace*{-5mm}
\caption{Energy of the residuals for the A1 and A2 events.}
\vspace*{-2mm}
\label{fig:A12_residuals_energy}
\end{figure}

The first scenario considers an oscillatory signal in the field voltage. To this end, the state variable $E_{fd}$ of the SEXS model is changed following the dynamics described in equation \eqref{eq:A1_EFd}. 
\begin{equation}
    E_{fd} = E_{fd_0} + a_{int_1} \sin \left( 2 \pi f_{int_1} \, t \right) \left[ 1 + a_{int_2} \sin \left( 2\, \pi 
    f_{int_2} \, t \right) \right]
    \label{eq:A1_EFd}
\end{equation}
The parameters used are: $E_{fd_0}=2.105$ p.u., $a_{int_1}=0.5$ p.u., $f_{int_1}=0.75$ Hz, $a_{int_2}= 0.1$ Hz and 
$f_{int_2}= 0.025$ Hz. The FO appears after 2s of simulation and remains for another 100s of simulation. The DSE is
performed for each generator following the procedure described in section \ref{sec:Method}, and three residuals are calculated for 
each step time. The energy of the residual signals with greater variability ($y_{max}$) are shown in Fig. 
\ref{fig:A12_residuals_energy}. As can be observed, the generator H is identified as the generator where the FO occurs 
since is the energy of the residuals is the lowest. However, the similarity of performing the DSE with generator G is 
highly marked in this case. 

\begin{figure*}[]
\centering
\vspace*{-2mm}
\includegraphics[width = 2\columnwidth]{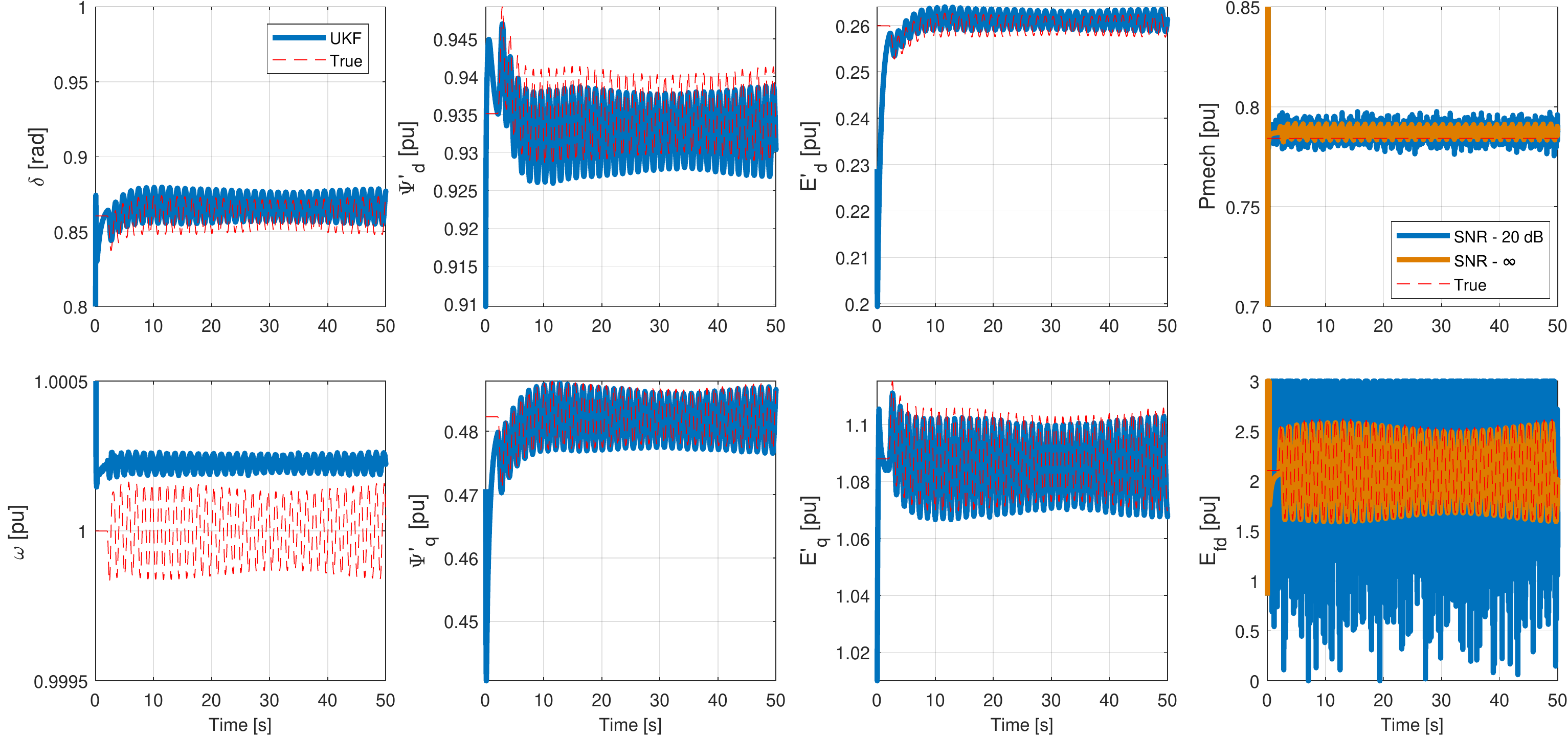}
%\vspace*{-5mm}
\caption{Generator states variables for the A1 event.}
\vspace*{-2mm}
\label{fig:A1_state_variables}
\end{figure*}

The estimated and true state variables of generator H are shown in Fig. \ref{fig:A1_state_variables}. It is concluded 
that the state variables can be estimated accurately. The greatest deviation between the estimation and the true states 
can be observed in the speed because an approximation was made in the measurement equations. Besides, it is interesting 
to note the effect of the AWGN in the estimation procedure. The estimations of the unknown inputs are degraded in a 
considerable way. For clarity reasons, the states for the $SNR= \infty$ case where not plotted in Fig. 
\ref{fig:A1_state_variables} because they are very similar to the case with the AWGN. 

\begin{figure}[]
\centering
\vspace*{-2mm}
\includegraphics[width = 1\columnwidth]{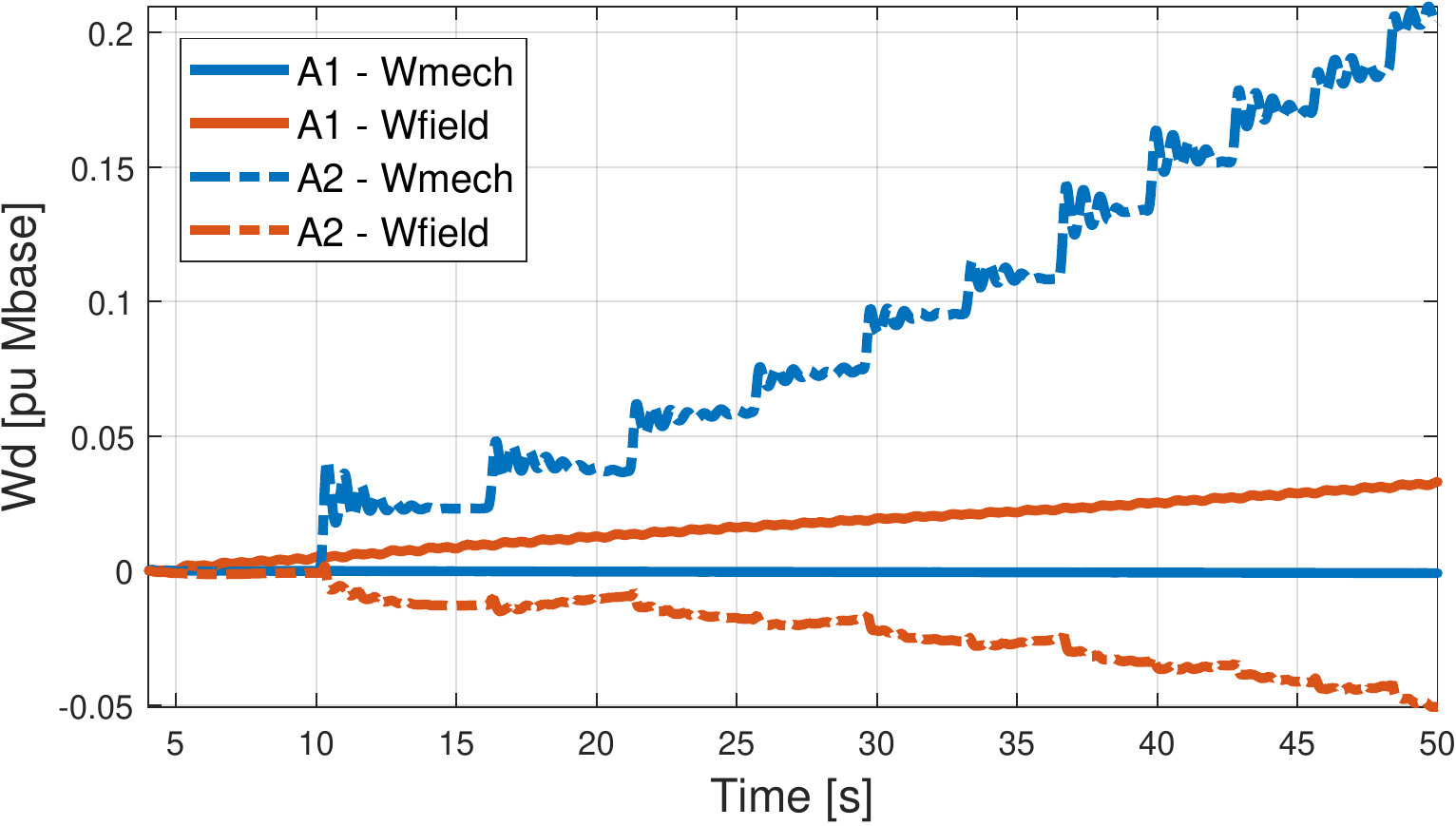}
%\vspace*{-5mm}
\caption{Energy functions for the test bench system.}
\vspace*{-2mm}
\label{fig:A12_energy_functions}
\end{figure}

Finally, the energy functions described in section \ref{sec:DEF} are calculated. The results are shown in \mbox{Fig. 
\ref{fig:A12_energy_functions}}. Here, the growing tendency for $W_{field}$ and the decreasing tendency for $W_{mech}$ 
enable the FO to be identified from the correct control loop.

\vspace{0.1cm}
\subsubsection{Perturbation on the mechanical power $P_{mech}$}
\begin{figure*}[]
\centering
\vspace*{-2mm}
\includegraphics[width = 2\columnwidth]{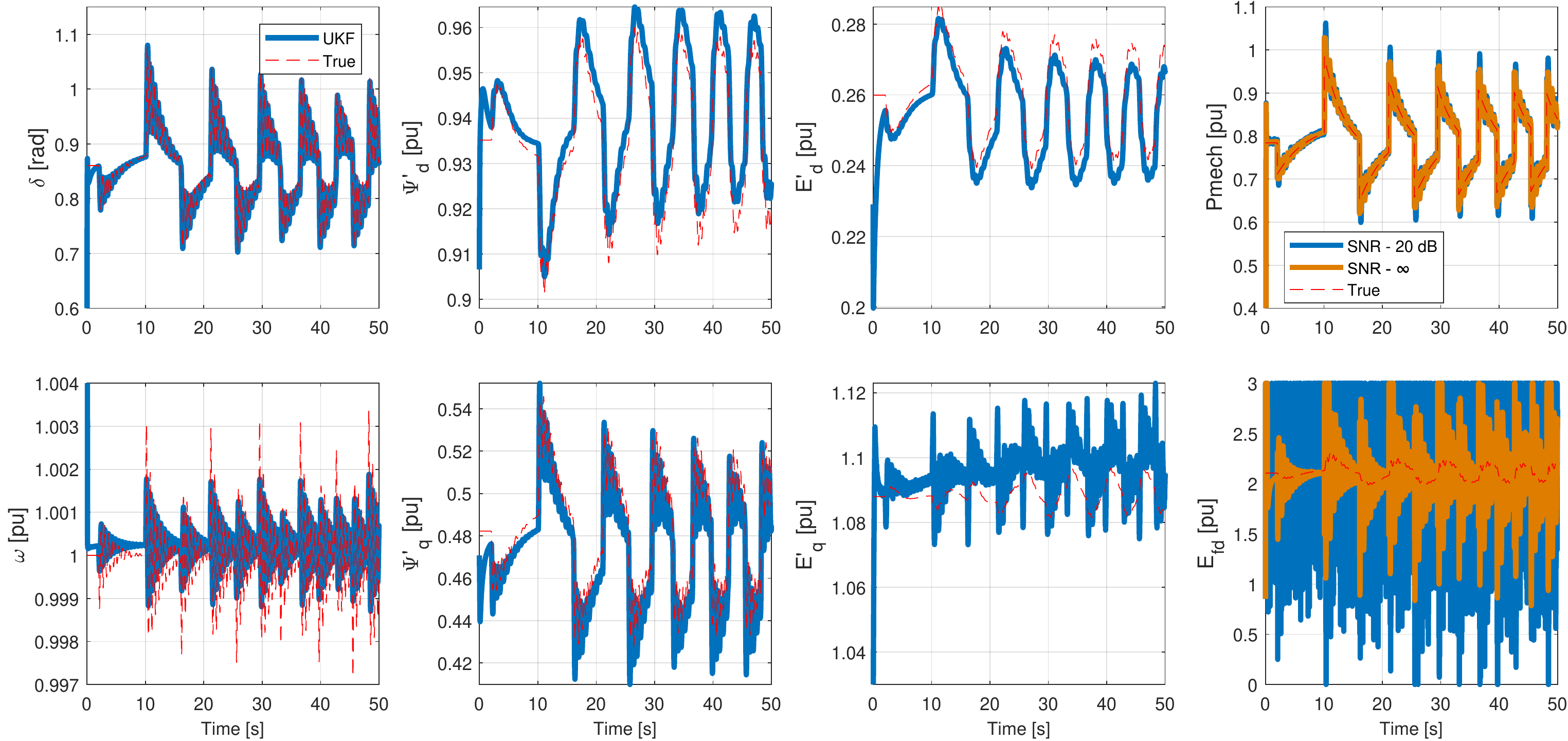}
%\vspace*{-5mm}
\caption{Generator states variables for the A2 event.}
\vspace*{-2mm}
\label{fig:A2_state_variables}
\end{figure*}

The second scenario to analyze consist in varying the control loop of the mechanical power. A non-stationary FO is simulated changing the gate signal ($g$) of the HYGOV model. We have chosen a square signal with a duty cycle equal to 0.5 and a linear time varying frequency $f_{g}(t)$: 

\begin{equation}
    g = \begin{cases}
    g_0\,(1+\frac{a_{g}}{2}), & \,\frac{K}{f_{g}(t)} \leq t < \frac{(K+1)}{2\,f_{g}(t)}\, \\
    g_0\,(1-\frac{a_{g}}{2}), &  \frac{(K+1)}{2f_{g}(t)}\, \leq t < \frac{(K+1)}{f_{g}(t)} 
    \end{cases}
\end{equation}

where $g_0=0.60278$, $a_{g}=0.1$ and $K=1,\cdots,5$. The initial value of $f_{g}$ is set to 0.05 and it reaches a maximum
of 0.2 at the time of 50s. After that the frequency decreases until the initial value at the time of 100s. In this case, 
the amplitude of the FO is higher than in the last scenario, causing more relevant variations in the output signals at
the point of connection. Hence, and it can be observed in Fig. \ref{fig:A12_residuals_energy}, for this event is easier 
to identify the generator H as the one where the FO take place. The mechanical power, and the other non-stationary 
states, are shown in Fig. \ref{fig:A2_state_variables}. It can be noticed that the DSE algorithm estimates the state
variables accurately, and the results are similar to the ones obtained in the section above. Again, the additive noise has a negative impact in the estimation procedure but the states can be estimated accurately enough. Finally, in Fig. \ref{fig:A12_energy_functions} there is a clear positive tendency for the function related to the $P_{mech}$ variable and a negative one for the $E_{fd}$ making it easy to identify a FO in the mechanical power control loop.

\subsection{IEEE-NASPI Oscillation Source Location Contest}

In this section, the raw data used in the IEEE-NASPI Oscillation Source Location Contest \cite{OSL} is used to test the 
proposed methodology. The simulated events are divided in cases where different scenarios are characterized. In this 
paper only 7 of 13 cases are analyzed. For example, an harmonic resonates with a natural mode of the system and produces 
the largest oscillation amplitude (case 12), the forced oscillation resonates with a regional inter-area mode (case 8), 
two forced oscillations are present (case 10). The other cases can be not analyzed with our proposal because some of the 
cases do not comply with the  requirement of having a PMU at the bus where the FO occurs, either the FO is originated 
from a load or HVDC \cite{OSLcases}. 

For each analyzed case, synthetic PMU measurements of bus voltage and branch current phasors, from the bus were the FO is
originated, are provided. For 30 s the system remains in a stationary state. After that, the event occurs during 60 s, so
a total of 90 s of data are used in each case. White Gaussian noise was added to the loads to represent random load 
fluctuations. Powertech Lab's TSAT software was used for the time-domain simulation. EPRI's PMU Emulator was used to 
process the simulation results to simulate PMU device performance, a mix of P Class (2-cycle window) and M Class (6-cycle
window) PMUs were used. EPRI's Synchrophasor Data Conditioning Tool was used to process the synthetic PMU data to 
introduce data quality problems. Each data set consisted of four text files for bus voltage magnitudes, bus voltage 
angles, branch current magnitudes, and branch current angles. Using the these data sets our proposal can be tested. 

\begin{figure}[]
\centering
\vspace*{-2mm}
\includegraphics[width = 1\columnwidth]{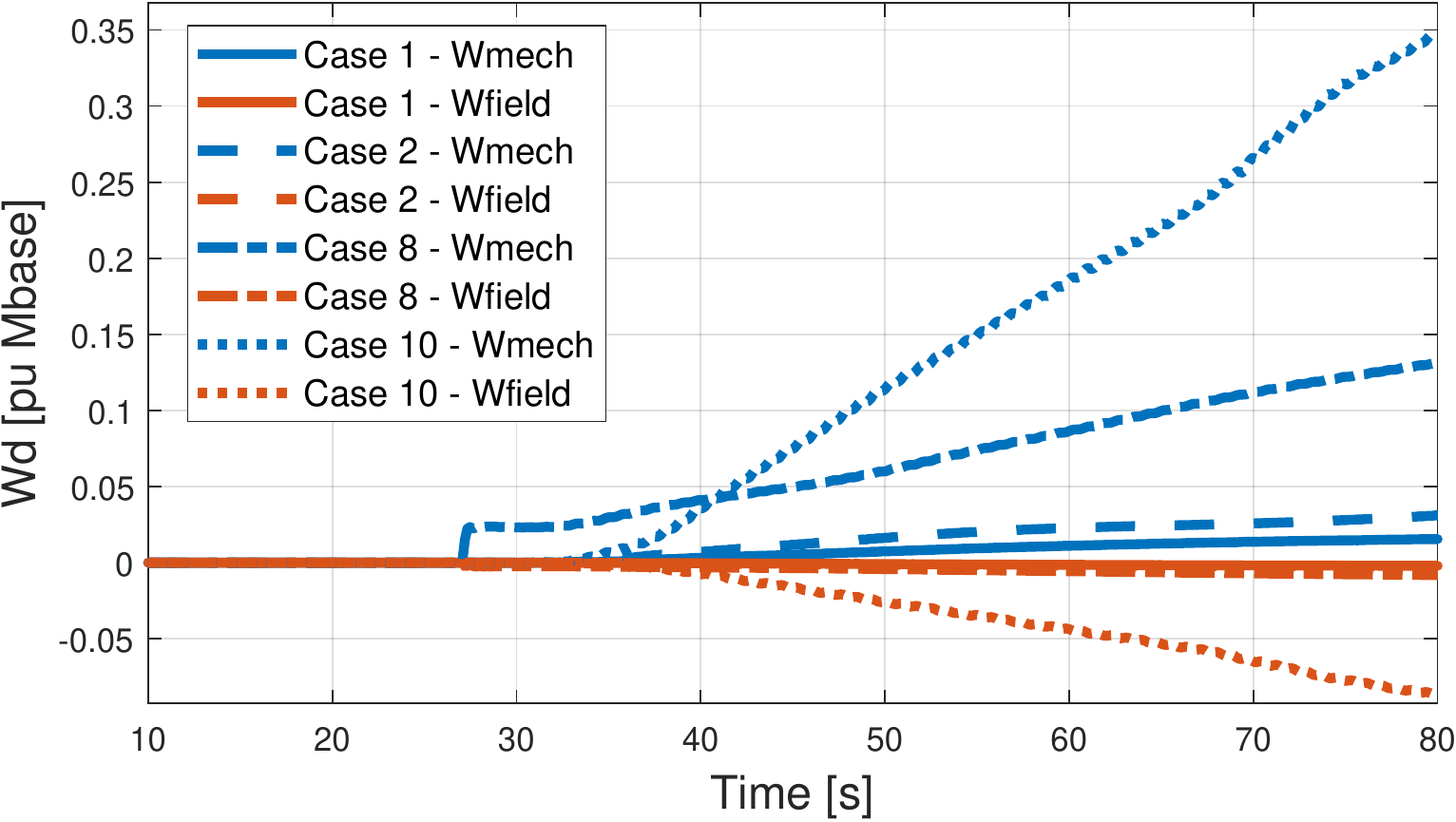}
%\vspace*{-5mm}
\caption{Energy functions for cases of the IEEE-NASPI oscillation source location contest.}
\vspace*{-2mm}
\label{fig:Naspi_energy_functions_part1}
\end{figure}

\begin{figure}[]
\centering
\vspace*{-2mm}
\includegraphics[width = 1\columnwidth]{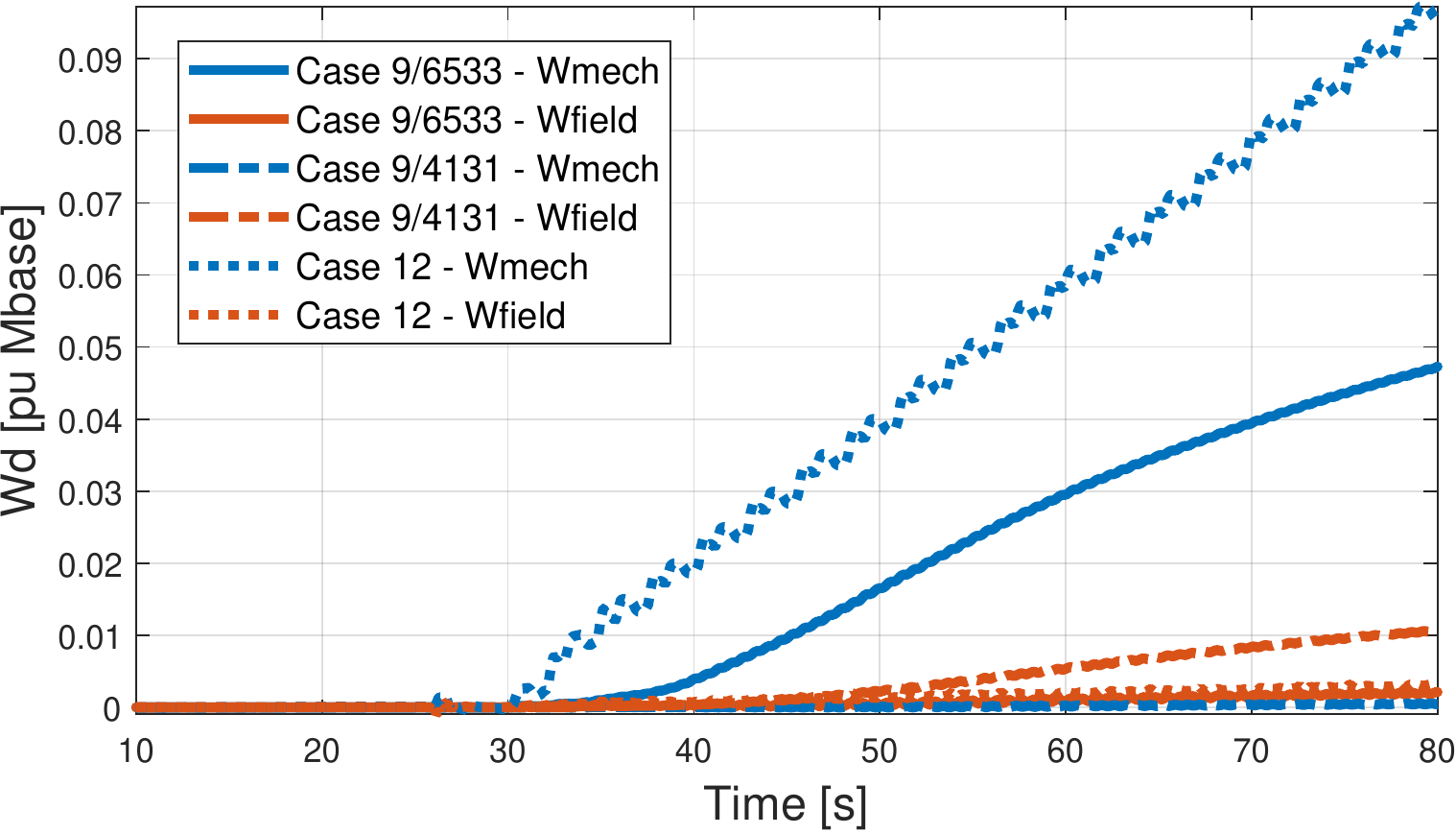}
%\vspace*{-5mm}
\caption{Energy functions for cases of the IEEE-NASPI oscillation source location contest.}
\vspace*{-2mm}
\label{fig:Naspi_energy_functions_part2}
\end{figure}

We present the results of applying the proposed methodology in Fig. \ref{fig:Naspi_energy_functions_part1} and Fig. 
\ref{fig:Naspi_energy_functions_part2}. Here, it can be seen that the cases 1, 2, 8 and 10 can be recognized clearly as 
cases where the FO take place in the mechanical power control loop. In these cases, the energy function of the 
$P_{mech}$ has the positive tendency while the energy of $E_{fd}$ has a negative one. For cases 9/6533 and 12 the energy 
function of the field control loop ends being one order of magnitude (at least) less than the value of the function of 
the mechanical power control loop. So if the requirements for the identification criterion are relaxed to include these 
types of scenarios the identification should be made correctly. Finally, Fig. 
\ref{fig:Naspi_energy_functions_part2} shows that the only case where the FO is detected from the field control 
loop is the 9/4131 case. We assume that that the energy function related to the control loop where the FO occurs does 
not have a negative tendency because the energy functions are defined for systems where the effects of the PSS are 
neglected. In the cases of Fig.\ref{fig:Naspi_energy_functions_part2} the damping is reduced by adjusting PSS gain 
in generator creating negative contribution into damping from that generator. Comparing these results with the solution 
key of the contest \cite{OSLsolutionkey}, it can be concluded that the FO sources, form all the analyzed cases, can be 
located correctly. It is worth to mention that from 21 participant teams, our team reached the third place 
\cite{OSLwinners}.

\section{Conclusions}

This paper proposes a systematic methodology for location forced oscillation sources when they are originated in one of the 
control loops of a synchronous generator. To achieve this goal we propose to combine the information obtained after a dynamic 
estimation process of each generator is performed, with a dissipating energy flow method. Besides, we have also shown a practical solution when a lack of observability is considered in the bus where the measurements are obtained. The core of this approach is based on a comparison between the performance of multiple DSE procedures. We have assume that only one FO is taking place in the bus system, and it can be identified because the event playback simulation generates the respective outputs of the other generators.    

The effectiveness of the proposed methodology was tested on simulated data, and also applied successfully in the cases of the 
IEEE-NASPI OSL Contest 2021. As we have shown the identification of the generator can be compromised when the energy of the oscillation is not strong enough to produce significant variations in the voltage and current at the connection point. Besides, some limitations can be observed when an AWGN was introduced. It can be concluded that if the TVE is not low enough, to guarantee a good signal to noise ratio between the FO and the noise, the performance can be highly degraded. Future works will be related to extend this methodology to include other observability issues. For example, to include scenarios where PMU measurements from a big area are the only information available.    

\appendices

\section{Nomenclature} \label{sec:nomenclature}
All the variables and constants used throughout this article are listed in Table \ref{tab:nomenclature}.
They are assessed using the per-unit system, unless they are time constants, which
are quantized in seconds, or phases, in radians.

\begin{table}[h!]
\caption{Nomenclature used for all variables and constants.}
\resizebox{1\columnwidth}{!}{
\centering
\normalsize
\begin{tabular}{|c|c|}
\hline
$\delta/\omega$ &     Rotor angle/angular velocity.  \tabularnewline \hline
$E'_{d}/E'_{q}$ &     d/q axis transient voltage.  \tabularnewline \hline
$\Psi'_{d}/\Psi'_{q}$ &     d/q axis transient flux.  \tabularnewline \hline
$P_{e}/E_{fd}$ &    Electric power / Field voltage.  \tabularnewline \hline
$P_{mech}/P_{m0}$ &   Instantaneous/Steady state mechanical power.  \tabularnewline \hline
%$I_{d}/I_{q}$ &    d/q axis stator current.  \tabularnewline \hline
%$R_{A}/X_{ls}$ &     Stator resistance/leakage reactance.  \tabularnewline \hline
$X_{ad}$ & Equivalent reactance from saturation function \tabularnewline \hline 
$X_{d},X{}_{q}$ &     d/q axis synchronous reactances.  \tabularnewline \hline
$X'_{d},X'_{q}$ &     d/q axis transient reactances.  \tabularnewline \hline
$X''_{d},X''_{q}$ &     d/q axis sub-transient reactances.  \tabularnewline \hline
$T'_{d},T'_{q}$ &     d/q axis transient open circuit time constants.  \tabularnewline \hline
$T''_{d},T''_{q}$ &     d/q axis sub-transient time constants.  \tabularnewline \hline
$D/H$ &     Damping factor / Inertia constant.  \tabularnewline \hline
\end{tabular} \label{tab:nomenclature}} 
\vspace*{.2cm}
\end{table}

\bibliographystyle{IEEEtran}

\bibliography{biblio}

\end{document}